\documentclass[aps,pre,twocolumn,groupedaddress]{revtex4}

\def\giorno{4/7/2005}

\def\*{{\bf ***}}

\def\a{\alpha}
\def\b{\beta}

\def\ga{\gamma}
\def\de{\delta}   
\def\eps{\varepsilon}
\def\phi{\varphi}

\def\s{\sigma}

\def\De{\Delta}

\def\pa{\partial}

\def\d{{\rm d}}       

\def\o+{\oplus}

\def\<{\langle}
\def\>{\rangle}

\def\interno{\hskip 2pt \vbox{\hbox{\vbox to .18
truecm{\vfill\hbox to .25 truecm
{\hfill\hfill}\vfill}\vrule}\hrule}\hskip 2 pt}

\def\({\left(}
\def\){\right)}
\def\[{\left[}
\def\]{\right]}
\def\=#1{\bar #1}
\def\~#1{\widetilde #1}
\def\.#1{\dot #1}
\def\^#1{\widehat #1}
\def\"#1{\ddot #1}

\def\ref#1{\cite{#1}}

\def\sec#1{ {}}

\begin{document}

\title{Asymptotic symmetries in an optical lattice}

\author{Giuseppe Gaeta}
\affiliation{Dipartimento di Matematica, Universit\`a di Milano,
via Saldini 50, I--20133 Milano (Italy)}
\email{gaeta@mat.unimi.it}

\date{\giorno }

\begin{abstract}
It was recently remarked by Lutz [{\it Phys. Rev. A} {\bf 67}
(2003), 051402(R)] that the equation for the marginal Wigner
distribution in an optical lattice admits a scale-free
distribution corresponding to Tsallis statistics. Here we show
that this distribution is invariant under an asymptotic symmetry
of the equation, hence that this scale-free behavior can be
understood in terms of symmetry analysis.
\end{abstract}


\maketitle

In a recent paper with R. Mancinelli \cite{RDS}
we have considered asymptotic symmetries of differential equations
and of their (asymptotic) solutions; in there we considered in
particular application to a given class of anomalous
reaction-diffusion equations which had been extensively studied
numerically \cite{MVV}, obtaining a theoretical explanation of the
observed long-time behavior of solutions. The same approach has
also been extended to a discrete version of these equations
\cite{GLM}.

In this note we apply our approach to a different kind of
anomalous diffusion; that is, we focus on the equation describing
anomalous transport in an optical lattice \cite{GR,Lut}. The
equation for the marginal Wigner distribution $w(p,t)$ of the
momentum $p$ at time $t$ reads
$$ w_t \ = \ - {\pa \over \pa p } \ \[ h(p) \, w \ - \ g(p) \, w_p \] \eqno(1) $$
where the functions $h(p)$ and $g(p)$ are given by
$$ h (p) \ := \ {- \a p \over 1 + (p / p_c)^2 } \ \ , \ \ g(p) \ := \ \ga_0 \ + \ {\ga_1  \over 1 + (p / p_c)^2 } \ ; \eqno(2) $$
here $\a , \ga_0 , \ga_1 $ are certain constants, $p_c$ represents the capture momentum.

Equation (1) should be complemented by a choice of the function
space to which the function $w(p,t)$ belongs. On physical grounds,
we require that $w(p,t)$ at fixed $t$ is normalizable and vanishes
asymptotically,
$$ |w(p,t)| \to 0 \ \ {\rm for} \ |p| \to \infty \ . \eqno(3) $$
We stress that there is a range of parameters ($- 1/2 < \mu $)
\cite{Lut} for which there is an equilibrium distribution with
these properties; this is the range to be studied here.

The explicit expressions of $h(p)$ and $g(p)$ make that, with the
shorthand notation $ \b (p ) := 1/[1 + (p/p_c)^2 ] $, the equation
(1), (2) reads
$$ \begin{array}{rl}
w_t \ =& \ \( \ga_0 + \b (p) \ga_1 \) \, w_{pp} \ + \\
 & \ + \ \( \a \b (p) \, - \, 2 \ga_1 [\b (p) / p_c ]^2 \) \, p \, w_p \ + \\
 & \ + \ \a \, \( \b (p) \, - \, 2 [ (p/p_c) \b (p) ]^2 \) \, w \ . \end{array} \eqno(4) $$

Attention to eq. (4) was recently called by Lutz \cite{Lut} (to
which the reader is also referred for derivation and a discussion
of this equation), who remarked that -- quite surprisingly -- the
equilibrium distribution is a Tsallis distribution, i.e. a
distribution optimizing the Tsallis entropy \cite{Tsa}. Indeed,
the stationary solution of (4) turns out to be
$$ w_0 (p) \ = \ {1 \over Z} \ \[ 1 \, - \, (\b  / \mu ) \, p^2 \]^\mu \eqno(5) $$
where
$$ \b  =  {\a \over 2 (\ga_0 + \ga_1 )} \ , \ \mu = {1 \over 1 - q} \ , \ q = 1 + {2 \ga_0 \over \a p_c^2} := 1 + \delta \ , \eqno(6) $$
and $Z$ is a normalization factor, which can be chosen so that
$\int_{- \infty}^{\infty} w_0 (p) \d p = 1 $. The physical range
is $q < 3$; the case $5/3 < q < 3$ (for which the second moment
$\int p^2 w_0 (p) \d p$ diverges) corresponds to anomalous moment
diffusion \cite{Lut}.

The peculiar properties of this equation and its stationary
solution, in particular concerning the power-law decay of $w_0
(p)$ for large $|p|$, hence its scale-free nature, has been
studied by Abe \cite{Abe} in connection with dilation symmetries
and canonical formalism.

Here we will consider generalized scaling transformations; that
is, local transformations acting as a standard scaling in the
independent $p$ and $t$ variables, and as a $p$ and $t$ dependent
scaling in the dependent variable $w(p,t)$. Moreover, we will not
require to have a full invariance of the equation, but be
satisfied in {\it asymptotic invariance} (for large $|p|$ and $t$
and hence, in view of (3), small $w$), as discussed in detail in
\cite{RDS}.

We start by recalling a
general feature of symmetry transformations and symmetry
invariance for PDEs and their solutions \cite{Gae,Olv,Win}; we
will specialize to a scalar PDEs for $w = w(p,t)$.

Under an infinitesimal transformation with generator
$$ X \ = \ \xi (p,t,w) \pa_p + \tau (p,t,w) \pa_t + \phi (p,t,w) \pa_w \eqno(7) $$
(that is, $p \mapsto \^p = p + \eps \xi (p,t,w)$, and so on) the
function $w(p,t)$ is mapped to a new function $\^w (p,t)$ with
$\^w (p,t) = w (p,t) + \eps [\de w(p,t)] $ and
$$ \de w \ = \ \[ \phi - w_p \xi - w_t  \tau \]_{w=w(p,t)} \ . \eqno(8) $$

Applying this to the distribution (5) and to generalized scalings
we get with easy computations that (up to a common factor) the
transformation generated by (7) with $\xi = - p $ leaves (5)
invariant if and only if
$$ \phi \ = \ (2 \, \b \, Z^{q-1} ) \, p^2 \, w^q \ ; \eqno(9) $$
obviously $\tau = \s t $ remains unrestricted at this stage.
Conversely, a function $w(p)$ is invariant under $X = - p \pa_p +
(2 \b Z^{q-1} ) p^2 w^q \pa_w$ if and only if $w(p)$ is of the
form (5); this is easily checked by using (8). We will write, from
now on,
$$ \de \ = \ q - 1 \ \ , \ \ \nu = 2 \b Z^\de \ . \eqno(10) $$

Albeit our analysis will be conducted at the infinitesimal level,
we note that the action of the one-parameter group (denote by $s$
the group parameter) generated by $ X = - p \pa_p + \nu p^2
w^{1+\de} \pa_w$ on the $p$ and $w$ variables is to map
$(p,w)=(p_0,w_0)$ into $(p(s) , w(s) )$ with
$$ \begin{array}{l}
p(s) \ = \ e^{- s} p_0 \ , \\
w(s) \ = \  \[ 1 + (\nu p_0^2/2) (1  - e^{- 2 s} ) w_0^\de
\]^{-(1/\de)} \ w_0 \ . \end{array} \eqno(11)
$$

We are now going to discuss what is
special in the generalized scaling vector fields identified above
$$ X \ = \ - p \pa_p \ + \ \s t \pa_t \ + \ \( \nu \, p^2 \, w^q \) \pa_w \eqno(12) $$
for what concerns their action on the equation (4).

We follow the general procedure for group analysis of differential
equations \cite{Gae,Olv,Win}. By acting with $X$ on the $(p,t,w)$
variables, we induce a transformation in the partial derivatives
of $w$ with respect to $p$ and $t$ as well (also called the second
prolongation of $X$); this is described -- restricting to partial
derivatives occurring in (4) -- by the vector field
$$ Y \ = \ X \ + \ \Psi_t {\pa \over \pa w_t} \ + \ \Psi_p {\pa \over \pa w_p} \ + \ \Psi_{pp} {\pa \over \pa w_{pp}} \ ; \eqno(13) $$
the coefficients $\Psi$ appearing in here is as follows: under the
infinitesimal transformation described by $X$, the partial
derivatives get transformed as $w_p \to w_p + \eps \Psi_p$, $w_t
\to w_t + \eps \Psi_t$, $w_{pp} \to w_{pp} + \eps \Psi_{pp}$.

When $X$ is the generalized scaling (12), i.e. for the mapping $p
\to (1-\eps ) p $, $t \to (1 + \eps \s ) t$, $ w \to (1 + \eps
(\nu p^2 w^\de )) w$,  the corresponding $\Psi$ can be easily
computed by the so-called prolongation formula \cite{Gae,Olv,Win}
to be
$$ \begin{array}{l}
\Psi_t \ = \ - \s w_t + \nu q p^2 w^\de \ ; \ \\
\Psi_p \ = \ w_p + 2 \nu p w^{(1+\de)} + \nu (1+\de) p^2 w^\de \ ; \\
\Psi_{pp} \ = \ 2 w_{pp} + 2 \nu w^{(1+\de)} + \\
\ \ \ \nu w^{(\de-1)} (1 + \de)
\[ \de p^2 w_p^2 + p w (4 w_p + p w_{pp} ) \]
\ . \end{array} \eqno(14) $$ These determine how the differential
equation (4) of interest here is transformed under $X$.

In computational terms, this is done by applying $Y$ on (4), and
substituting for $w_t$ according to (4) itself; the obtained
expression must be zero for $X$ to be a symmetry of the equation.
Proceeding in this way, we obtain a condition which we write in
compact form (see below for the explicit expressions of the
functions $A_k$) as
$$ A_0 (p,t,w) \ + \ A_1 (p,t,w) \, w_p \ + \ A_2 (p,t,w) \, w_{pp} \ = \ 0 \ . \eqno(15) $$
The functions $A_k$ must vanish separately for (15) to be
satisfied.

The explicit expression of $A_2$ is
$$ A_2 \ = \ - \ga_0 (2 + \s)
- {\ga_1 p_c^2 ( p_c^2 ( 2 + \s )  + ( 4 + \s ) p^2 ) \over [p_c^2
+ p^2 ]^2 } \ . \eqno(16) $$ The limit of this for large $|p|$ is
nonzero unless we choose, as we do from now on,
$$ \s \ = \ - 2 \ . \eqno(17) $$
With this choice, $X$ reads
$$ X \ = \ - p \, \pa_p \ - \ 2 \, t \, \pa_t \ + \ \nu \, p^2 \, w^q \, \pa_w \eqno(18) $$
and $A_2$ reduces to
$$ A_2 \ = \ {\frac{-2\,\ga_1\,{{p_c}^2}\,{p^2}}
   {{{\left( {{p_c}^2} + {p^2} \right) }^2}}} \ ; \eqno(19) $$
therefore, $A_2 \to 0$ for $|p| \to \infty$.

As for $A_1$, with the choice (17) it reads
$$ \begin{array}{l} A_1 \ = \
\[ 2 p \( ( p_c^2 + p^2 )
( \a p_c^4 - 2 ( 1 + \delta ) \ga_0 \nu w^\delta
( p_c^2 + w^2 )^2 ) \right. \right. \\
\left. \left. - 2 \ga_1 p_c^2 ( ( 1 + \delta )  \nu p_c^4 w^\delta
+ p^2 ( -1 + \nu w^\delta p^2 +
\delta \nu w^\delta p^2 ) \right. \right. \\
\left. \left. + p_c^2 ( 1 + 2 \nu w^\delta p^2 + 2 \delta \nu
w^\delta p^2 )  )  \) \] \, \times \, \[ p_c^2 + p^2
\]^{-3} \ . \end{array} \eqno(20) $$ This is a more involved
expression, but it is still easy to check that in the limit $|p|
\to \infty$ and $w \to 0$, recall (3), we get $A_1 \to 0$.

Finally, and again with (17), $A_0$ reads
$$ \begin{array}{rl}
A_0 \ =& \
\[ w ( -2 \nu w^\de
( p_c^2 + p^2 ) ( \ga_1 p_c^2 ( p_c^2 - p^2 ) + \right. \\
& \left. + \ga_0 ( p_c^2 +
p^2 )^2 )  + \right. \\
 & \left. + \a p_c^2
( - ( ( 2 + \delta )  \nu w^\de p^6 )  - 2 p_c^2 p^2
( 3 + 2 \nu w^\de p^2 )  + \right. \\
 & \left. + p_c^4
( 2 - 2 \nu w^\de p^2 + \delta \nu w^\de p^2 ) ) ) \] \, \times \\
 & \times \, \[ p_c^2 + p^2 \]^{-3} \ . \end{array} \eqno(21) $$ Once again,
for $|p| \to \infty$ and $w \to 0$ we get $A_0 \to 0$.

This computation shows on the one hand that the one-parameter
group of transformations generated by (18) is not a symmetry of
the equation (4), as the $A_k$ are not identically zero; on the
other hand, it shows that (18) is an asymptotic symmetry for this
equation, defined in the function space identified by (3). No
vector field of the form (12) with $\s \not= -2$ is an asymptotic
symmetry of (4); hence (18) is the only invariance generator for
(5) which is also an asymptotic symmetry of (4).

Let us discuss the effect on (18) on the
normalization condition (3). Using (8) we obtain that a generic
function $w(p,t)$ is mapped by (18) into $\^w = w + \eps \de w$
with $\de w = \nu p^2 w^q + p w_p + 2 t w_t$; on solutions to (1)
this reads
$$ \de w \ = \ \nu \, p^2 \, w^q \ + \ p \, w_p \ - \ 2 t \, {\pa \over \pa p } (h w - g w_p ) \ . \eqno (22) $$
With $ I [w] := \int_{- \infty}^{+ \infty} w(p,t) \d p$, we
obviously have $I[\^w] = I[w] + \eps I[\de w]$. In view of (22),
the latter amounts to
$$ I [\de w] \ = \ \nu I [p^2 w^q] \, + \, I[p w_p ] \, + \, 2 t
\, I[\pa_p (h w - g w_p ) ] \ . \eqno(23) $$ It is immediate to
see, using also an integration by parts for the last one, that (3)
implies the finiteness of the last two integrals in (23).

As for $I[p^2 w^q ]$, note that (3) holds if, for $|p| \to
\infty$, $w (p) \simeq 1/p^{2 k}$ with $k> 1/2$. For general
solutions $w(p,t)$, the condition of normalizable variation $kq >
3/2$ is more restrictive than the normalization condition $k >
1/2$ for all $q$ in the physical range $q < 3$.

Note that for $w (p)$ given by (5), $k = 1/\de$ and the condition
$kq>3/2$ holds for all $q<3$; that is, functions which are near to
the stationary solution $w_0 (p)$ will always have normalizable
variation under (18).

Thus, strictly speaking, our method should be applied only on
functions satisfying $|I[p^2 w^q ] | < \infty$, i.e. decaying (for
$|p| \to \infty$) faster than $1/|p|^{3/q}$. As observed above,
this includes all functions near to the Tsallis distribution (5),
for all $q$ in the physical range.

The (asymptotic) invariance of (4) under
(18) could also be analyzed using the systematic procedure of
\cite{RDS}; this requires to introduce symmetry-adapted
coordinates $(v,y,\s)$, with $v$ the dependent variable. These are
the coordinates in which (18) is simply $ X = - 2 \s \pa_\s$, and
its action on derivatives up to order two is described by $ Y = X
+ 2 v_\s (\pa / \pa v_\s )$ \cite{Gae,Olv,Win}. In the present
case, adapted coordinates are $\s = t$, $ y = p^2/t $, $v = w^{-
\de } - (\nu \de / 2) p^2$.

Finally, let us briefly mention differences
with the work by Abe \cite{Abe}. Abe considered a dilation
symmetry of (4), based on an auxiliary field $\Lambda (p)$ which
could be not normalizable. The Abe symmetry has generator $ G =
\int p \Lambda_p \, w \ \d p $. In the language of symmetry theory
\cite{Gae,Olv,Win} this is a nonlocal symmetry (it depends on an
integral over $p$ rather than just on the value of $w$ at the
point $p$), whereas here we considered local ones. Moreover Abe
worked in canonical formalism, which requires to select a
symplectic structure, whereas here we did not consider any
additional structure. The two approaches are thus quite different;
it appears that symmetries considered by Abe are more general,
while the one considered here has the advantage of being a {\it
local} one.

\medskip
Let us summarize our discussion. We have
considered the equation for the marginal Wigner distribution
$w(p,t)$ of the momentum $p$ at time $t$ in an optical lattice
from the point of view of symmetry analysis. It is known that this
equation admits the distribution $w_0 (p)$ as a stationary
solution.

We have identified the generalized scaling invariance group of
$w_0 (p)$, described by (11) and generated by $X = - p \pa_p + (2
\b Z^{q-1} p^2 w^q ) \pa_w$.

We passed then to consider generalized scalings (12) acting on the
time variable as well; it is easily seen that these cannot be
symmetries of the equation (4). We investigated then if they may
be {\it asymptotic symmetries} for that equation, and observed
that a necessary condition for this is provided by (17). With this
choice, it turns out that we have indeed an asymptotic symmetry,
whose generator is (18), of our equation.

This shows that the Tsallis distribution (4) is an invariant
function for a transformation $X$ which is an asymptotic symmetry
(for large $|p|$) of the equation under study, and hence that {\it
the scale-free asymptotic behavior of $w_0 (p)$ is a consequence
of the asymptotic symmetry properties of the equation (4)}
describing the characteristics of the optical lattice.

\medskip
I gratefully thank the referees for constructive remarks, and R.
Mancinelli for useful discussions. This work was partially
supported by {\it GNFM-INdAM} under the project ``Simmetria e
tecniche di riduzione''.

\end{document}